\documentclass[prb,fleqn,12pt,showpacs]{revtex4}
\usepackage{epsfig}
\usepackage{graphicx}
\usepackage{amsfonts}
\begin{document}
\title{Spin waves in the $(0,\pi)$ and $(0,\pi,\pi)$ ordered \\ 
SDW states of the $t-t'$ Hubbard model: \\ 
Application to doped iron pnictides}
\author{Nimisha Raghuvanshi and Avinash Singh}
\email{avinas@iitk.ac.in}
\affiliation{Department of Physics, Indian Institute of Technology Kanpur}
\begin{abstract}
Spin waves in $(0,\pi)$ and $(0,\pi,\pi)$ ordered spin-density-wave (SDW) states of the $t-t'$ Hubbard model are investigated at finite doping. In the presence of small $t'$, these composite ferro-antiferromagnetic (F-AF) states are found to be strongly stabilized at finite hole doping due to enhanced carrier-induced ferromagnetic spin couplings as in metallic ferromagnets. Anisotropic spin-wave velocities, spin-wave energy scale of around 200meV, reduced magnetic moment, and rapid suppression of magnetic order with electron doping $x$ (corresponding to F substitution of O atoms in $\rm La O_{1-x} F_x Fe As$ or Ni substitution of Fe atoms in $\rm Ba Fe_{2-x} Ni_x As_2$) obtained in this model are in agreement with observed magnetic properties of doped iron pnictides.  
\end{abstract}
\maketitle
\newpage

\section{Introduction}
Following the recent discovery of superconductivity\cite{kamihara_2008,takahashi_2008,chen_2008,rotter_2008} 
in doped iron pnictides such as $\rm R O_{1-x} F_x Fe As$ (R = La, Ce, Nd, Sm, Gd), there has been a strong interest in their magnetic properties as well. Single-crystal neutron scattering studies of their parent compounds have indicated a commensurate magnetic ordering with iron moments ordered ferromagnetically in the $b$ direction and antiferromagnetically in the $a$ and $c$ directions.\cite{goldman_2008} Inelastic neutron scattering measurements in $\rm A Fe_2 As_2$ (A = Ca, Ba, Sr) yield sharp spin-wave excitations on an energy scale $\sim 200$meV.\cite{zhao_2008,zhao_2009,diallo_2009}

All known compounds in these classes contain FeAs layers and exhibit a common phase diagram, with parent compounds exhibiting a magnetically ordered SDW state below $T_N \approx 200$ K, and onset of superconductivity following the suppression of long-range magnetic order by electron doping or pressure. Contrast with cuprates has naturally followed in order to gain insight from the significant progress made in understanding superconductivity and magnetism in correlated electron systems.\cite{cuprate_review} While onset of superconductivity at finite doping is a common feature, in contrast to the Mott insulating nature of cuprates, the pnictides appear to be commensurate SDW systems in the intermediate coupling regime.
Appreciable hybridisation between Fe 3d orbitals and As 3p orbitals possibly reduces the effective correlation term $U$ as compared to cuprates.\cite{norman_2008} Comparison with X-ray photoemission spectra of the calculated density of states (DOS) within the LDA + DMFT approach supports the physical picture of a multi-band metal with intermediate correlations.\cite{anisimov_2008,aichhorn_2009}

In cuprates, the intense interest in the nature of magnetic excitations in the quantum antiferromagnet, their coupling with hole motion and scrambling of local AF order, strongly incoherent hole spectral function, and possibility of pairing interaction mediated by exchange of local magnetic excitations have contributed significantly to obtaining insight and understanding of their magnetic and electronic properties. 

Therefore, an investigation of magnetic excitations in the $(0,\pi)$ and $(0,\pi,\pi)$ ordered SDW states within a minimal itinerant electron model should be of interest, particularly within an approach which is valid in the full range of interaction strength including the relevant weak and intermediate coupling regimes. In this paper, we will therefore consider the $t-t'$ Hubbard model, and obtain the magnon energies in the F-AF states in two and three dimensions, focussing especially on the role of finite $t'$ and doping in stabilizing the F-AF order. 

Spin waves in the $(0,\pi)$ and $(0,\pi,\pi)$ ordered states were investigated earlier in the undoped $t-t'$ Hubbard model,\cite{phase} and have been recently investigated within $J_1 - J_2$ and $J_{1a} - J_{1b} - J_2$ Heisenberg models on a square lattice.\cite{yao_2008,applegate_2010} Reduced magnetic moments and suppression of magnetic ordering have been investigated in terms of associated magnetic frustration effect in such models.\cite{si_2008,manousakis_2010}
Spin excitations in SDW states have been investigated within two-band models involving excitonic instability of nested electron-like and hole-like Fermi pockets.\cite{brydon_2009} Doping dependence of spin fluctuations and electron correlations have been theoretically investigated within an effective five band Hubbard model for iron pnictides using the FLEX approximation.\cite{ikeda_arxiv_2010} 


\section{The $t-t'$ Hubbard model}
We consider the $t-t'$ Hubbard model on square and simple cubic lattices, with hopping terms $t$ and $t'$ between nearest-neighbour (NN) and next-nearest-neighbour (NNN) pairs of sites, respectively:
\begin{equation}
H = -t \sum_{i,\delta, \sigma} ^{\rm NN} a_{i, \sigma}^{\dagger} a_{i+\delta, \sigma}
-t' \sum_{i,\kappa, \sigma} ^{\rm NNN} a_{i, \sigma}^{\dagger} a_{i+\kappa, \sigma}
+  U\sum_{i} n_{i \uparrow} n_{i \downarrow} \; .
\end{equation}
The observed asymmetry in the antiferromagnetism of cuprates with respect to hole and electron doping,\cite{chubukov_95,asingh_2002} existence of metallic antiferromagnetism at half filling,\cite{duffy_97,hofstetter_98} and correlated motion of electrons in metallic ferromagnets in fcc and bcc lattices,\cite{ulmke_1998,vollhardt_2001,nolting_book,singh_2006,pandey_2007} exemplify few physical systems which have been investigated in terms of the above model.

The two-sublattice representation of the AF state (corresponding to ordering wave-vector $(\pi,\pi)$ in two dimensions) conveniently allows for investigation of spin waves, quantum corrections, N\'{e}el temperature, hole dynamics etc. In analogy, we will consider F-AF SDW states with ${\bf Q}=(0,\pi)$ and $(0,\pi,\pi)$ involving ferromagnetic spin ordering in one direction and antiferromagnetic spin ordering in the remaining direction(s). The self-consistent-field (Hartree-Fock) approximation provides a convenient basis in which many-body correlations effects can then be systematically incorporated. The two-sublattice structure for the F-AF SDW states in two and three dimensions are given below. 

\subsection{${\bf Q}=(0,\pi)$ ordered state}
In this state, spins are ordered ferromagnetically along the x direction and antiferromagnetically along the y direction. The NN hopping terms in the x direction connect sites of same sublattice, while NN hopping terms in the y direction and NNN hopping terms connect sites of opposite sublattices. The HF Hamiltonian matrix in the two-sublattice basis therefore takes the form:
\begin{equation}
H_{\rm HF}^\sigma ({\bf k}) = \left [ \begin{array}{cc} -\sigma \Delta + \epsilon_{\bf k} ^x  \;\;
& \epsilon_{\bf k} ^y +  {\epsilon'_{\bf k}} \\ \epsilon_{\bf k} ^{y}  + {\epsilon'_{\bf k}} \;\;
& \sigma \Delta + \epsilon_{\bf k} ^x  \end{array} \right ] 
\equiv \eta'_{\bf k} \; {\bf 1} + \left [ \begin{array}{cc} -\sigma \Delta & \eta_{\bf k} \\ 
\eta_{\bf k} & \sigma \Delta  \end{array} \right ]
\end{equation}
for electron spin $\sigma$, where $\eta'_{\bf k} \equiv  \epsilon_{\bf k} ^x = -2t\cos k_x $ and $\eta_{\bf k} \equiv  \epsilon_{\bf k} ^{y} + {\epsilon'_{\bf k}} = -2t\cos k_y - 4t' \cos k_x \cos k_y$, and the self-consistent exchange field is given by $2\Delta=mU$ in terms of the sublattice magnetization $m$.

The SDW state quasiparticle band energies: 
\begin{equation}
E_{{\bf k}\sigma}^{(\pm)} = \eta'_{\bf k} \pm \sqrt{\Delta^2 + \eta_{\bf k} ^2}
\end{equation}
corresponding to the two spins ($\sigma=\uparrow,\downarrow$) and the two $(\pm)$ bands, and the quasiparticle amplitudes $a_{{\bf k}\sigma}$ and $b_{{\bf k}\sigma}$ on the two sublattices A and B are given by:
\begin{eqnarray}
a_{{\bf k}\uparrow\ominus}^2 = b_{{\bf k}\downarrow\ominus}^2 = a_{{\bf k}\downarrow\oplus}^2 =
b_{{\bf k}\uparrow\oplus}^2 &=& \frac{1}{2}\left ( 1+ \frac{\Delta}{\sqrt{\Delta^2 +\eta_{\bf k} ^2}} \right ) \nonumber \\
a_{{\bf k}\uparrow\oplus}^2 = b_{{\bf k}\downarrow\oplus}^2 = a_{{\bf k}\downarrow\ominus}^2 =
b_{{\bf k}\uparrow\ominus}^2 &=& \frac{1}{2}\left ( 1- \frac{\Delta}{\sqrt{\Delta^2 +\eta_{\bf k} ^2}} \right ).
\end{eqnarray}
These relationships follow from the spin-sublattice and particle-hole symmetry in the AF state. The above two expressions provide the majority and minority fermionic densities. On the A-sublattice, the majority density is of spin $\uparrow$ ($\downarrow$) states in the lower (upper) band.

\subsection{${\bf Q}=(0,\pi,\pi)$ ordered state}
In this state, spins are ferromagnetically ordered along the x directions and antiferromagnetically ordered along the y and z directions. The NN hopping terms in the y-z plane connect sites of opposite sublattices, while those in the x direction connect sites of the same sublattice. Similarly, NNN hopping terms in the y-z plane connect sites of the same sublattice while those in the z-x and x-y planes connect sites of opposite sublattices. The HF Hamiltonian matrix therefore takes the form:
\begin{equation}
H_{\rm HF}^\sigma ({\bf k}) = \left [ \begin{array}{cc}
-\sigma \Delta + \epsilon_{\bf k} ^x + {\epsilon'_{\bf k}}^{yz} \;\;
& \epsilon_{\bf k} ^{y} + \epsilon_{\bf k} ^{z} +  {\epsilon'_{\bf k}}^{zx} + {\epsilon'_{\bf k}}^{xy} \\
\ \\ 
\epsilon_{\bf k} ^{y} + \epsilon_{\bf k} ^{z} + {\epsilon'_{\bf k}}^{zx} + {\epsilon'_{\bf k}}^{xy} \;\;
& \sigma \Delta + \epsilon_{\bf k} ^x + {\epsilon'_{\bf k}}^{yz} \end{array} \right ]  
\equiv \eta'_{\bf k} \; {\bf 1} +  \left [ \begin{array}{cc}
-\sigma \Delta & \eta_{\bf k}  \\ \eta_{\bf k} & \sigma \Delta  \end{array} \right ]
\end{equation}
where
\begin{eqnarray}
\eta'_{\bf k} & \equiv & \epsilon_{\bf k} ^x + {\epsilon'_{\bf k}}^{yz}
=-2t\cos k_x - 4t' \cos k_y \cos k_z   \\ 
\eta_{\bf k} & \equiv & \epsilon_{\bf k} ^{y} + \epsilon_{\bf k} ^{z} + 
{\epsilon'_{\bf k}}^{zx} + {\epsilon'_{\bf k}}^{xy} = [-2t-4t' \cos k_x ] (\cos k_y + \cos k_z) \nonumber
\end{eqnarray}
The quasiparticle band energies and amplitudes are again as in Eqs. (3,4). 

\section{Magnon propagator}
Magnon excitations in the F-AF state, corresponding to transverse spin fluctuations about the ordering direction (assumed z), are obtained from the time-ordered propagator of the transverse spin operators $S_i ^-$ and $S_j ^+$ at sites $i$ and $j$:
\begin{equation}
\chi^{-+}({\bf q},\omega) = \int dt \sum_i 
e^{i\omega(t-t')} e^{-i{\bf q}.({\bf r}_i -{\bf r}_j)} \; 
\langle \Psi_{\rm G} | T [ S_i ^- (t) S_j ^+ (t')]|\Psi_{\rm G}\rangle \; .
\end{equation}
In the random phase approximation (RPA), the magnon propagator:
\begin{equation}
\chi^{-+} _{\rm RPA} ({\bf q},\omega) = \frac{[\chi^0 ({\bf q},\omega)]}{{\bf 1} -U[\chi^0 ({\bf q},\omega)]} 
\end{equation}
in terms of the bare particle-hole propagator $[\chi^0({\bf q},\omega)]$, evaluated by integrating out the fermions in the spontaneously-broken-symmetry F-AF state. In the insulating state, involving only interband particle-hole processes, the bare propagator is given by:
\begin{eqnarray}
[\chi^0({\bf q},\omega)] &=& \sum_{\bf k}
\left [ \begin{array}{lr} 
a_{{\bf k}\uparrow \ominus}^2  a_{{\bf k-q}\downarrow \oplus}^2  & 
a_{{\bf k}\uparrow \ominus}b_{{\bf k}\uparrow \ominus}a_{{\bf k-q}\downarrow \oplus}b_{{\bf k-q}\downarrow \oplus} \\
a_{{\bf k}\uparrow \ominus}b_{{\bf k}\uparrow \ominus}a_{{\bf k-q}\downarrow \oplus}b_{{\bf k-q}\downarrow \oplus} &
b_{{\bf k}\uparrow \ominus}^2  b_{{\bf k-q}\downarrow \oplus}^2  \end{array}
\right ] 
\frac{1}{E_{{\bf k-q}}^{\oplus} - E_{\bf k} ^{\ominus} + \omega -i \eta}
\nonumber \\
&+& \sum_{\bf k}
\left [ \begin{array}{lr} 
a_{{\bf k}\uparrow \oplus}^2  a_{{\bf k-q}\downarrow \ominus}^2  & 
a_{{\bf k}\uparrow \oplus}b_{{\bf k}\uparrow \oplus}a_{{\bf k-q}\downarrow \ominus}b_{{\bf k-q}\downarrow \ominus} \\
a_{{\bf k}\uparrow \oplus}b_{{\bf k}\uparrow \oplus}a_{{\bf k-q}\downarrow \ominus}b_{{\bf k-q}\downarrow \ominus} &
b_{{\bf k}\uparrow \oplus}^2  b_{{\bf k-q}\downarrow \ominus}^2  \end{array}
\right ] 
\frac{1}{E_{{\bf k}}^{\oplus} - E_{{\bf k-q}} ^{\ominus} - \omega -i \eta}
\end{eqnarray}
in terms of the quasiparticle amplitudes and energies. In the antiferromagnetic metallic (AFM) state, additional intraband processes involving particle-hole excitations from the same band also contribute.\cite{asingh_2002,af_met} Evaluation of $[\chi^0 ({\bf q},\omega)]$ and the RPA-level magnon propagator $\chi^{-+}({\bf q},\omega)$ in the strong coupling limit is described in the next section. 

\section{Insulating $(0,\pi,\pi)$ state in the strong coupling limit}
In this section we consider the analytically simple strong coupling limit, and evaluate the magnon propagator in the insulating $(0,\pi,\pi)$ state at half filling, illustrating the competition between NN ($J=4t^2/U$) and NNN ($J'=4t'^2/U$) AF spin couplings and consequent instability of this state as $J'$ drops below $J/4$. 

In the strong coupling limit, the majority and minority quasiparticle densities in Eq. (4) reduce to approximately $1-\eta_{\bf k} ^2/4\Delta^2$ and $\eta_{\bf k} ^2/4\Delta^2$, respectively. Similarly expanding the energy denominators involving the quasiparticle band energies in powers of $t/\Delta$, $t'/\Delta$, and $\omega/\Delta$, and systematically retaining terms up to order $t^2/\Delta^2$ and $t^{'2}/\Delta^2$ in the bare particle-hole propagator, we obtain for the RPA level magnon propagator:\cite{phase}
\begin{equation}
[\chi^{-+}({\bf q},\omega)] = -\frac{1}{2} \left (\frac{2J}{\omega_{\bf q}} \right ) 
\left [ \begin{array}{lr} {\cal A}_{\bf q} -\frac{\omega}{2J} & {\cal B}_{\bf q} \\
{\cal B}_{\bf q} & {\cal A}_{\bf q} + \frac{\omega}{2J} \end{array} \right ] 
\left ( \frac{1}{\omega-\omega_{\bf q} + i \eta} - \frac{1}{\omega + \omega_{\bf q} -i\eta} \right )
\end{equation} 
where the magnon propagator matrix elements:
\begin{eqnarray}
{\cal A}_{\bf q} &=& \left ( 1 + \frac{2J'}{J}\right ) -\frac{1}{2} \left \{ (1 - \cos q_x) + \frac{2J'}{J}(1-\cos q_y \cos q_z) \right \} \nonumber \\
{\cal B}_{\bf q} &=& \left (1 + \frac{2J'}{J}\cos q_x \right ) \frac{1}{2} (\cos q_y + \cos q_z)
\end{eqnarray} 
\noindent
and the magnon-mode energies are obtained as:
\begin{eqnarray}
\left ( \frac{\omega_{\bf q}}{2J} \right )^2 = {\cal A}_{\bf q} ^2 - {\cal B}_{\bf q} ^2 
&=& \left [ \left ( 1+ \frac{2J'}{J} \right ) -\frac{1}{2}
\left \{(1-\cos q_x) + \frac{2J'}{J}(1-\cos q_y \cos q_z)\right \} \right ]^2 \nonumber \\
&-& \left [ \left ( 1+ \frac{2J'}{J}\cos q_x \right ) \left (\frac{\cos q_y +\cos q_z}{2}\right ) \right ]^2
\end{eqnarray}

In the long wavelength limit, the magnon energy reduces to
\begin{equation}
\left ( \frac{\omega_{\bf q}}{2J} \right )^2 \approx \frac{1}{2}\left ( 1+\frac{2J'}{J}\right )
[ \alpha  q_x^2 + q_y ^2 +q_z^2 ] \; ,
\end{equation}
where the coefficient $\alpha = \left ( \frac{4J'}{J} -1 \right )$ of the $q_x^2$ term becomes negative for $J'/J < 1/4 $, signaling the instability of the F-AF phase at $J'/J = 1/4 $. Anisotropic spin wave velocities naturally follow from the different $q^2$ coefficients in the F and AF directions.

The above instability can also be seen from energy considerations. The classical energy per spin for the two orderings are: $E_{\rm AF}=-6J+12J'$ and $E_{\rm F-AF}=-2J-4J'$, so that the F-AF state becomes energetically favourable for $J' > J/4$. In three dimensions, the colinear ${\bf Q}=(0,\pi,\pi)$ state is stable even at the classical level, unlike the degeneracy present in the $d=2$ case at this level. 

As an illustration of quantum corrections beyond the HF approximation, the spin-fluctuation correction $\delta m_{\rm SF}$ to the sublattice magnetization in the F-AF phase, which can be evaluated from the magnon propagator in terms of the transverse spin correlations,\cite{phase} is shown in Fig. 1. The correction in the AF phase is also shown for comparison. Near the transition point $J'/J=1/4$, the correction in the F-AF phase is seen to be nearly half of that in the AF phase, indicating greater robustness of the F-AF phase with respect to quantum spin fluctuations. The spin-fluctuation correction in both phases approaches 1 (the HF value of sublattice magnetization) only very close to the critical value $J'/J=1/4$. This implies that (up to first order) $m$ vanishes only very close to $J'/J=1/4$, so that the extent of the spin-disordered phase is quite narrow. This is unlike the $d=2$ case, where the AF order is lost at $J'/J \approx 0.37$, well before the F-AF state appears at $J'/J \gtrsim 0.5$.  

\begin{figure}
\vspace*{-0mm}
\hspace*{0mm}
\psfig{figure=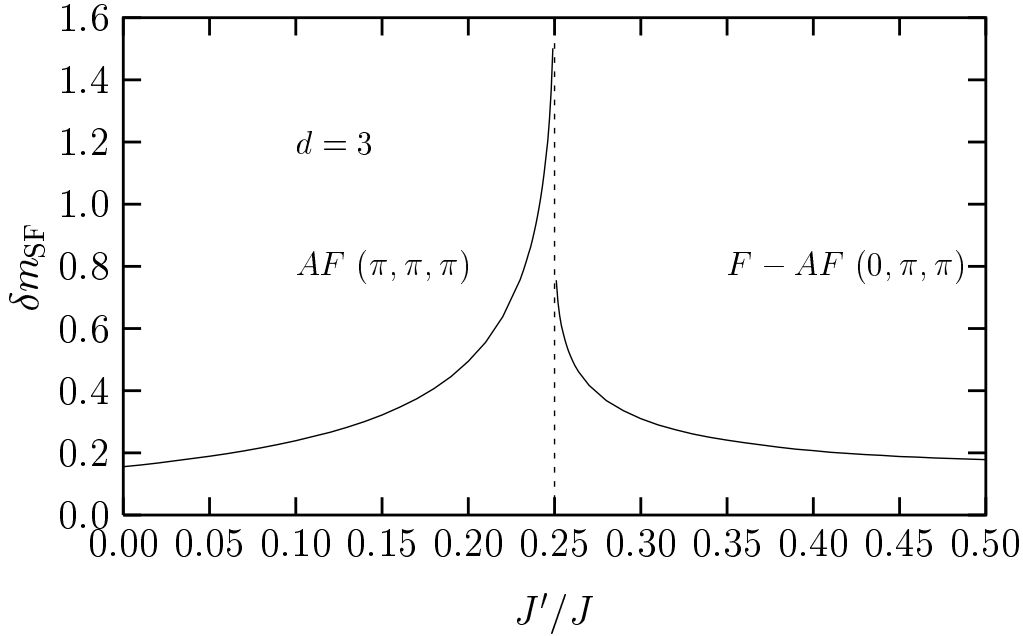,width=135mm,angle=0}
\vspace*{-0mm}
\caption{The spin-fluctuation correction to sublattice magnetization in the AF and the F-AF phases.}
\end{figure}

\section{Stabilization of the hole-doped F-AF state}
As shown by the strong-coupling analysis of the previous section, AF NNN spin couplings $J'=4t'^2/U$ generated by $t'$ stabilize the undoped F-AF state (for $t'/t > 1/\sqrt{2}$ in two and $t'/t > 1/2$ in three dimensions). So how is the F-AF state affected by other physical processes which might also generate effective NNN spin couplings? 

AF NNN spin couplings are generated even in the absence of $t'$, as in the $t-U$ model at finite doping. In fact, it is these effectively frustrating spin couplings, generated by intra-band particle-hole processes, which are responsible for destabilizing the AF state at any finite doping, as observed in hole doped cuprates.\cite{asingh_2002} However, as the F-AF state is actually stabilized rather than being frustrated by the AF NNN spin couplings, these doping-induced spin couplings should actually favour the F-AF state by supplementing the $t'$-induced spin couplings. 

More importantly, carrier-induced F NN spin couplings responsible for metallic ferromagnetism become increasingly important at finite doping. This is especially so in presence of small $t'$, which can cause strongly peaked electronic spectral distribution due to band dispersion saddle points $({\mbox{\boldmath $\nabla$}} \epsilon_{\bf k} = 0)$, which strongly enhance band ferromagnetism by increasing the delocalization contribution $\sim \langle {\mbox{\boldmath $\nabla$}} ^2 \epsilon_{\bf k} \rangle$ to spin stiffness while strongly suppressing the correlation-induced exchange contribution $\sim \langle ({\mbox{\boldmath $\nabla$}} \epsilon_{\bf k})^2/U \rangle$ due to correlated motion of electrons.\cite{singh_2006,pandey_2007} The F-AF state at finite doping is therefore expected to be stabilized at even lower $t'$ values. In this section we will show that indeed finite hole doping and small $t'$ strongly stabilizes the F-AF state.


Fig. 2 show the spin wave energy along symmetry directions in the Brillouin zone for the doped F-AF state, with orderings $(0,\pi)$ and $(0,\pi,\pi)$ as considered earlier. The variation of wave-vector ${\bf q}$ follows the sequence $(0,0)\rightarrow (0,\pi)\rightarrow (\pi,\pi)\rightarrow (0,0)\rightarrow (\pi,0)\rightarrow (\pi,\pi)$ in $2d$
and $(0,0,0)\rightarrow (0,\pi,\pi)\rightarrow (0,\pi,0)\rightarrow (\pi,\pi,0)\rightarrow (\pi,0,0)\rightarrow (0,0,0)$ in $3d$. In order to focus on doping dependence, the SDW bands were kept unchanged with fixed $\Delta/t=4$. The rapid crossover from negative to positive energy magnon modes shows a strong stabilization of the F-AF state upon hole doping. This stabilization occurs for much smaller $t'$ values compared to the critical values required in the undoped F-AF state. A finite $t'$ is quite realistic in view of the hybridization between Fe and As orbitals. 

The anisotropic spin wave velocities, evident from the different slopes in the F and AF directions in Fig. 2(a), can be understood readily in terms of the independent origin (delocalization and exchange) of the effective F and AF spin couplings, resulting in different coefficients of $q_x ^2$ and $q_y ^2$, as in Eq. (13). Furthermore, for $t\sim 200$meV and $U$ in the intermediate coupling regime, the calculated spin wave energy scale of around 200meV is as observed in neutron scattering measurements of iron pnictides. 

Are the doping-induced spin couplings sufficient to stabilize the F-AF state without any finite $t'$? We find that in the absence of $t'$ the F-AF state is not stabilized for any doping. While AF and F orderings do get separately stabilized at low and high hole dopings, respectively, as indicated by the spin-wave dispersion along AF and F directions, both are not simultaneously stabilized. Thus, small $t'$ plays a crucial role in stabilizing the doped F-AF state with respect to transverse spin fluctuations in the entire Brillouin zone. A possible link between orthorhombic distortion and an effective NNN hopping $t'$ would then explain why this distortion appears necessary for magnetic ordering to be stabilized in iron pnictides.

\begin{figure}
\vspace*{-0mm}
\hspace*{0mm}
\psfig{figure=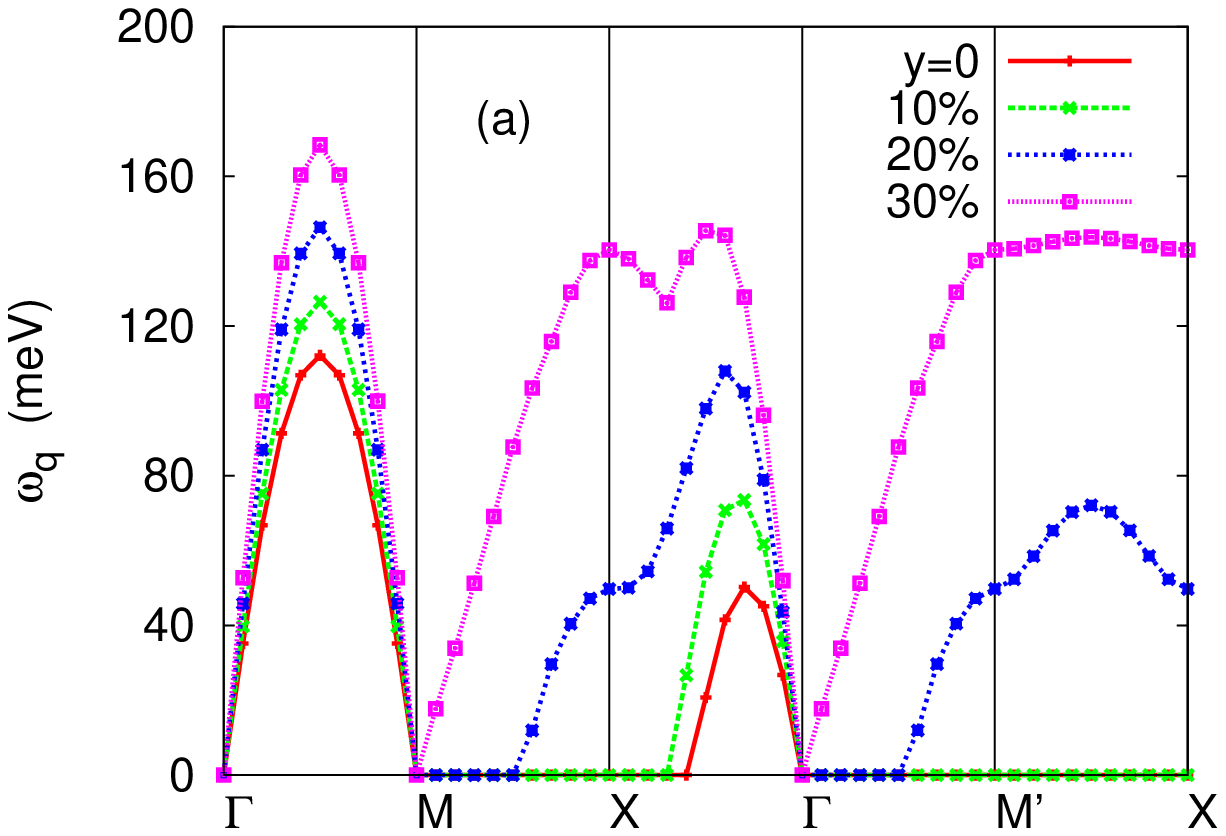,width=80mm,angle=0}
\psfig{figure=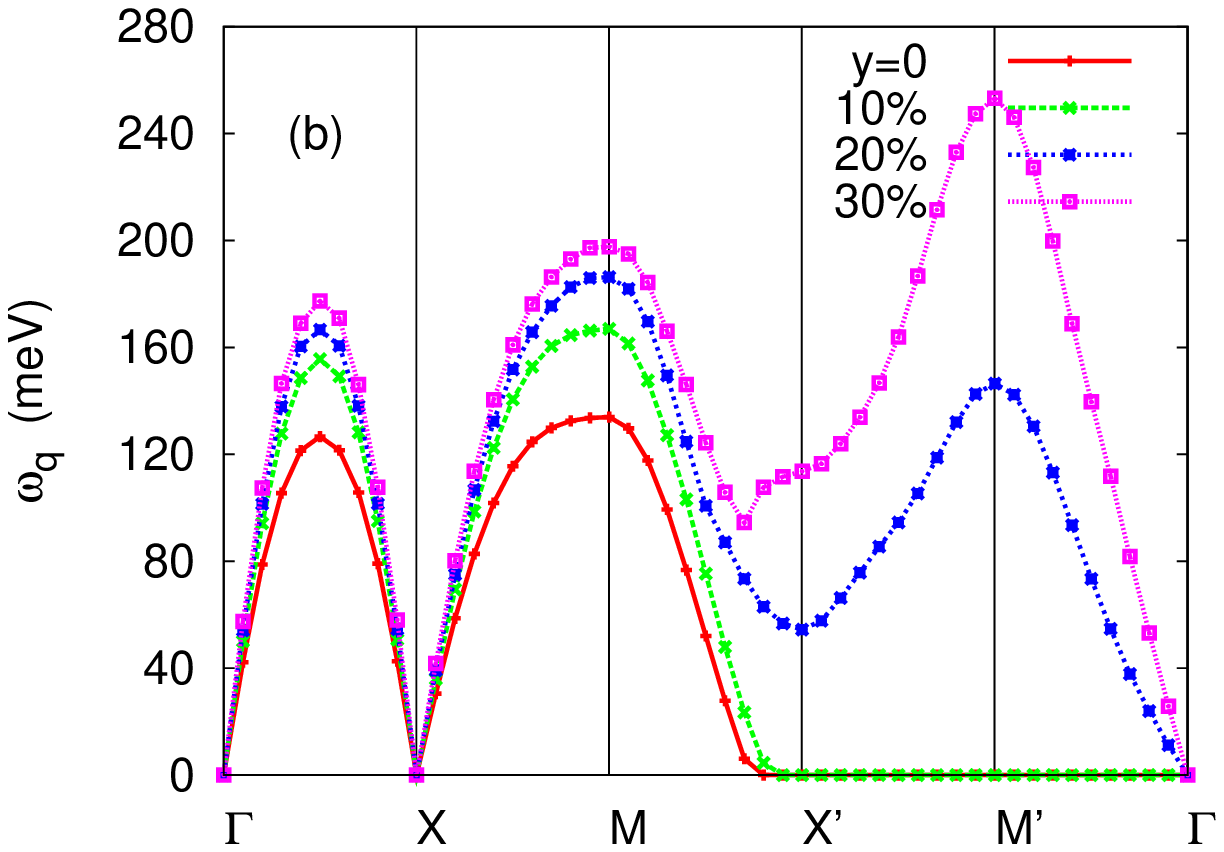,width=80mm,angle=0}
\vspace*{-0mm}
\caption{Spin wave energy along symmetry directions in the Brillouin zone for the F-AF states with (a) $(\pi,0)$ and (b) $(0,\pi,\pi)$ orderings. The rapid crossover from negative to positive energy modes, especially in the F direction $\Gamma$-M', shows a strong stabilization of the F-AF states upon hole doping $y$. Here $t'/t=0.5$ in (a) and 0.3 in (b).}
\end{figure}

As the maximal sensitivity of magnon modes to doping is found along the ferromagnetic direction ($\Gamma$-M'), we have examined their doping dependence at an intermediate wavevector ${\bf q}=(\pi/2,0,0)$. A pre-doped level with hole doping $y_0$ was taken to represent the partially-filled band of the parent compound, with electron doping $x=y_0 - y$ defined as the reduction in hole doping from this level (representing $\rm F^-$ substitution of $\rm O^{2-}$ in the doped pnictides). Fig. 3 shows the behaviour of the magnon mode energy with electron doping. The rapid magnon energy suppression and crossover to negative energy modes indicating destabilization of the F-AF state provides an understanding of the observed rapid suppression of magnetic ordering temperature in iron pnictides with electron doping (due to F substitution of O atoms in $\rm La O_{1-x} F_x Fe As$ or Ni substitution of Fe atoms in $\rm Ba Fe_{2-x} Ni_x As_2$).\cite{wang_2010} Here, the interaction strength was fixed at $U=8t$ in the intermediate coupling regime, and $y_0 = 0.35$.

The observed reduced magnetic moment in pnictides can be understood in terms of depletion of the predominantly magnetic states. Characteristic of the SDW state, electronic states at the top of the band are significantly more magnetic than states deeper in the band, especially in the weak-coupling limit. Therefore hole doping of these states rapidly diminishes the local magnetic moment $m=2\Delta/U$, which is reduced to $\approx 0.4$ at $x=0$ in Fig. 3. While local moments will get enhanced on filling up these empty magnetic states by electron doping (F substitution of O), the rapidly diminished carrier-induced F NN spin couplings are then unable to sustain the F-AF ordering. This highlights the two distinctly different mechanisms behind the observed reduced magnetic moment and the suppression of magnetic ordering in iron pnictides. Enhanced spin-fluctuation quantum correction in the vicinity of the magnetic instability point (as in Fig. 1) is also possibly important in reducing the magnetic moment.

\begin{figure}
\vspace*{-0mm}
\hspace*{0mm}
\psfig{figure=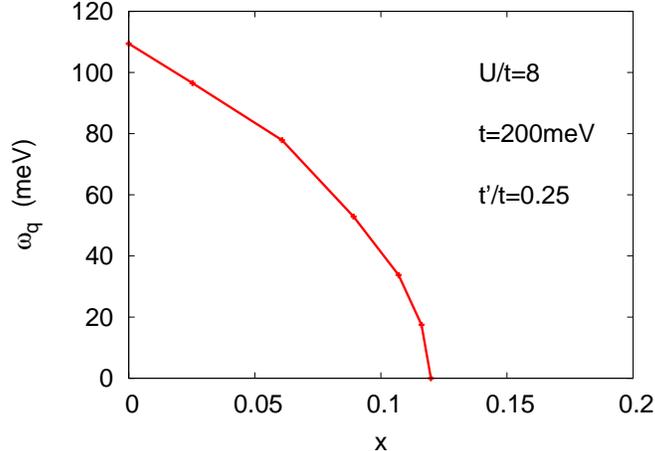,width=90mm,angle=0}
\vspace*{-0mm}
\caption{Electron doping ($x$) dependence of the magnon energy at wavevector ${\bf q}=(\pi/2,0,0)$, showing rapid suppression of magnon mode energy along the ferromagnetic ordering direction and destabilization of the F-AF state.}
\end{figure}

\section{Conclusions}
Spin waves in the $(0,\pi)$ and $(0,\pi,\pi)$ ordered F-AF SDW states of the $t-t'$ Hubbard model were investigated at finite doping in the intermediate coupling regime. Including both inter-band and intra-band processes in the particle-hole propagator to obtain the spin-wave propagator which incorporates effective spin couplings, spin-wave dispersion was obtained along different symmetry directions in the Brillouin zone. The F-AF state was found to be strongly stabilized at finite hole doping, as evidenced by the rapid crossover from negative to positive spin-wave energies, most noticeably in the F direction. 

This stabilization was ascribed mainly to carrier-induced ferromagnetic spin couplings (along the F chains) as in metallic ferromagnets, and was found to be strongly enhanced in presence of finite $t'$ due to band structure saddle point effect. This calculated doping behaviour is in agreement with the observed rapid suppression of the magnetic order in doped pnictides such as $\rm La O_{1-x} F_x Fe As$ on electron doping arising from F substitution of O atoms. The doping and $t'$ dependence of spin waves obtained within the minimal $t-t'$ Hubbard model accounts for many of the observed magnetic properties of iron pnictides, including anisotropic spin wave velocities, spin wave energy scale, reduced magnetic moment, and rapid suppresion of magnetic ordering with electron doping. The spin-wave velocity anisotropy is strongly enhanced on electron doping near the instability point. 

Quantum corrections to spin waves beyond RPA should be of interest in order to incorporate the correlated motion of electrons, as  investigated in the metallic ferromagnetic state recently,\cite{singh_2006,pandey_2007} and the insulating AF state earlier,\cite{quantum} incorporating correlation-induced self-energy and vertex corrections within a systematic inverse-degeneracy $(1/{\cal N})$ expansion scheme which preserves spin-rotation symmetry and hence the Goldstone mode at each order. 

In an orbitally degenerate metallic ferromagnet, the inter-orbital Coulomb interaction (Hund's coupling) $J$ was recently shown to strongly suppress quantum corrections,\cite{hunds} particularly for large orbital degeneracy ${\cal N}$, with the quantum correction magnitude   determined by an effective quantum parameter $\frac{[1+({\cal N}-1)J/U]^2} {[1+({\cal N}-1)(J/U)^2]}$. The renormalized spin stiffness with realistic parameters was obtained in close agreement with the measured spin stiffness for bcc Fe. In iron pnictides, the reduced effective ${\cal N}$ due to hybridization between (magnetic) Fe and (non-magnetic) As orbitals should enhance quantum corrections and hence suppress (F-direction) spin wave energies and the ordering temperature. These considerations are relevant in view of recent magnetic form factor studies indicating that multiple d orbitals of iron atoms are occupied.\cite{ratcliff_2010}

The magnon energy scale also determines finite temperature magnetic properties. Magnon thermal excitation yields the fall off of magnetization with temperature and, within a self-consistently renormalized spin fluctuation theory, the magnetic ordering temperature. Thus, the rapid suppression of magnon energy with electron doping obtained in this paper accounts for the observed reduction of magnetic ordering temperature on F substitution of O atoms in $\rm La O_{1-x} F_x FeAs$. 
Furthermore, spatially anisotropic magnetic couplings, with small ratio $(J_\perp / J_\parallel = r \ll 1)$ of interlayer to planar magnetic couplings, reduces the ordering temperature to $\sim T_c ^{\rm iso} /\ln (1/r)$ compared to the isotropic case, which possibly accounts for the low ordering temperature observed in doped pnictides ($\sim 200$K), as in the layered cuprate antiferromagnet $\rm La_2 Cu O_4$ where $J\sim 1500$ K and magnon energy scale $\sim 250$meV, but the N\'{e}el temperature is only about 400K.

\end{document}